\documentclass{article}[]
\usepackage[affil-it]{authblk}
\usepackage{mathtools}
\usepackage{amsthm}
\usepackage{amssymb}
\usepackage{gensymb}
\usepackage{wasysym}
\usepackage{dsfont}
\usepackage{graphicx}
\usepackage{tabu}
\usepackage{longtable}
\usepackage{makeidx}
\usepackage{multicol}
\usepackage{listings}
\usepackage{caption}
\usepackage{subcaption}
\usepackage{wrapfig}
\usepackage{cancel}
\usepackage{url}

\makeindex

\begin{document}

\title{Using social network graph analysis for interest detection}
\author{Brian Lee Yung Rowe
\thanks{Electronic address: \texttt{rowe@zatonovo.com}}}
\affil{Zato Novo, LLC}
\affil{Data Analytics, School of Professional Studies, CUNY}

\makeatletter
\def\@maketitle{%
  \newpage
  \null
  \vskip 2em%
  \begin{center}%
  \let \footnote \thanks
    {\Large\bfseries \@title \par}%
    \vskip 1.5em%
    {\normalsize
      \lineskip .5em%
      \begin{tabular}[t]{c}%
        \@author
      \end{tabular}\par}%
    \vskip 1em%
    {\normalsize \@date}%
  \end{center}%
  \par
  \vskip 1.5em}
\makeatother
\maketitle
\setcounter{page}{1}

\begin{abstract}
A person's interests exist as an internal state and are difficult to define.
Since only external actions are observable, a proxy must be used that
represents someone's interests. Techniques like collaborative filtering,
behavioral targeting, and hashtag analysis implicitly model an
individual's interests. I argue that these models are limited to shallow,
temporary interests, which do not reflect people's deeper interests or 
passions.
I propose an alternative model of interests that takes advantage of
a user's social graph. The basic principle is that people only follow
those that interest them, so the social graph is an effective and robust proxy 
for people's interests.
\end{abstract}

\section{Introduction}
Taken together, the Internet and social media have ushered in an 
era of mass personalization. 
Businesses can now target advertisements based on 
behavioral tracking and generate product recommendations 
based on surfing behavior.
To many, these technologies finally fulfill the longstanding 
mantra of \emph{know thy customer}.
Proponents of these approaches trumpet the fact that anonymity is preserved,
thus creating an ideal situation for both businesses and consumers.
However, knowing your customer -- or more broadly, your audience --
is about relationships and not just targeted marketing.
Building relationships is about connecting with people at 
an individual level with a message that is both authentic and 
that resonates with each person.
To do so requires more than displaying a travel advertisement 
after someone visits a travel site.
It's about understanding people’s passions and connecting with them 
in an authentic way.
This approach builds loyalty and intimacy between a brand and its audience.
At Zato Novo, we've developed quantitative models that identify
any Twitter user's interests to facilitate authentic outreach. 

Let's see how authentic outreach benefits everyone.
Suppose you are a non-profit organization and are holding a fundraiser.
As part of the fundraiser you want to have businesses sponsor a raffle.
What types of items should you raffle off? 
By knowing your supporters' interests, 
you can make a data-driven decision rather than leaving it to chance.
This is better for everyone involved: 
volunteers and supporters get products they want,
businesses expose their products and services to a relevant audience,
and non-profits draw a bigger turnout.
A similar process is applicable to political campaigns or 
any event organizer that has corporate sponsors. 

Matchmaking is another area of business that can take 
advantage of individual interests. 
Dating sites can use interests to suggest activities 
that are compatible with prospective couples.
Non-profits can identify the areas of expertise or influence of 
their volunteers, granting them operational scale 
that may have gone untapped.
Businesses can similarly identify brand ambassadors 
among their followers and leverage them to 
connect with audiences previously out of reach. 
This works because people's passions have a permanence to them,
whereas interests derived from browsing behavior is more ephemeral.

Social media is an obvious choice for authentic outreach.
Hashtag campaigns are a marketing tool designed to
generate buzz and engagement around an idea or cause 
related to a brand. Doing so not only raises awareness 
around the organization running the campaign but 
can also offer insights into the people engaging with the campaign.
These campaigns can be optimized for a specific audience 
by analyzing audience interests.
To start, a well-planned hashtag campaign defines the desired audience.
The audience can be drawn from similar campaigns.
This stage is a good time to identify social media amplifiers
(active users with many followers).
Optimal messaging is then crafted based on the interests of the audience.
Similarly, ads promoting the campaign can be optimally targeted 
based on audience interests.
Tapping into your brand amplifiers and ambassadors further spreads the message.
As the campaign progresses, analyzing the interests of the people
engaged with the campaign enables organizations to verify the 
effectiveness of the campaign.

\section{Interests Defined}
Leveraging people's interests for authentic outreach is straightforward.
It is less straightforward defining an individual's interests.
Unlike the physical sciences,
there is no equation that defines human interests precisely.
We can really only know someone's interests by 
externally observing their behavior. 
\footnote{including what they've written or spoken about} 
This hints at the need to find a proxy for interests with 
some measurable phenomenon.
One approach is to observe people's actions in relation to content or products.
The assumption is that a positive review indicates a good product 
while a negative review indicates a poor quality product.
A similar assumption drives the assumption that buying a product 
indicates interest in that product or product category.
For instance, buying a camera likely indicates an interest in photography.
But if the camera was bought as a gift, then the conclusion would be false.
Similarly, a single camera purchase cannot tell you 
how avid a photographer someone is and if she would be likely
to purchase multiple cameras in a year.

\subsection{Collaborative filtering}
\begin{figure}
\centering
\begin{subfigure}[b]{0.45\textwidth}
\includegraphics[width=0.9\textwidth]{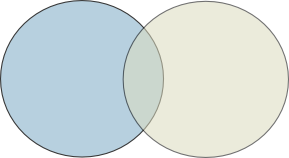}
\caption{A small intersection yields less accurate predictions}
\end{subfigure}
\begin{subfigure}[b]{0.45\textwidth}
\includegraphics[width=0.9\textwidth]{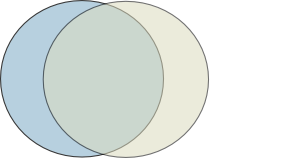}
\caption{A large overlap yields greater accuracy}
\label{fig:large-intersection}
\end{subfigure}
\caption{Two users with shared interests}
\label{fig:collab-filter}
\end{figure}

A single individual's purchasing or browsing behavior 
is often not enough information to predict future buying behavior,
let alone their underlying interests. 
However, it is possible to extrapolate a user's interests 
based on common browsing and buying behavior of all users on a given platform.
This technique is known as collaborative filtering
\cite{Schafer07collaborativefiltering}
and is motivated by observing how friends recommend
products to each other. For example, suppose you have a friend that
is a film critic. 
She has recommended numerous films to you
and you've liked all of them. Given this overlap in taste,
you'll probably like her future recommendations as well.
This relationship can be described using basic set theory:
given two users $a$ and $b$ with observed interests
$A$ and $B$ and common interests $A \cap B \neq \emptyset $, 
we say that $A' = B - A$ is a recommendation for user $A$ and
$B' = A - B$ is a recommendation for user $B$.
As the relative size of $A \cap B$ grows vis a vis the size of $A$ and $B$,
you would expect that the accuracy of the recommendations would improve.
Figure \ref{fig:collab-filter} illustrates the difference in these scenarios.
At scale it is easy to imagine that the more in common someone has with others,
the more accurate the recommendations will be for the non-overlapping behavior. 

As compelling as this theory is,
collaborative filtering is only appropriate for certain situations.
Returning to the movie example, why are our friend's recommendations
so good? The answer is that she knows our tastes. We may only like
Westerns, while our friend watches Westerns, Science Fiction, Film Noir,
among others. Implicit in her recommendations is the filtering
out of other movies that she likes that aren't Westerns.
A naive collaborative filtering algorithm would have no way of
partitioning her recommendations based on the genre of the movie
\footnote{Semantic tagging and other methods can make collaborative filtering
more robust}.

Here's another example.
Consider the following scenario about two men:
Bill lives in a house in the suburbs while Bob lives in an apartment.
Suppose these two men own the same car, the same home theater system,
and have the same taste in music, movies, and clothes.
Hence their interests overlap to a high degree as seen in 
Figure \ref{fig:large-intersection}.
One day Bill buys a lawn mower and rates it 5 stars.
Since Bill and Bob share so many interests,
should Bob be recommended a lawn mower?

Collaborative filtering often relies on users rating products,
and it is not always clear why a rating is given.
A movie may garner a 5 star rating due to great acting,
stunning visual effects, great cinematography, etc. 
On the flip side, a 1 star rating may be motivated by poor resolution, 
bad dubbing or subtitles, a bad ending, and so on.
Ratings, therefore, act as a dimensional reduction technique.
Since a single rating encodes many unstructured and variable factors,
ratings have the perverse effect of introducing noise as opposed to
removing noise from the inference process.

\subsection{Behavioral targeting}
Collaborative filtering is usually site-specific and 
therefore not applicable across websites. 
It is also difficult for businesses with a narrow focus 
to leverage collaborative filtering.
After all, how many stereo systems does one person need?
For broader targeting, businesses rely on ad networks that 
track a user's surfing behavior across websites \cite{behavioraltargeting}.
The basic assumption is that users only visit websites they are interested in.
Therefore content and shopping sites can indicate a user's interests.
Continuing the photography example, if a user is active on Flickr or 
National Geographic and also a camera review site, 
then presumably the user has an interest in photography.
Once a user leaves these sites and begins browsing other webpages
the ad network can deliver advertisements related to cameras.
Targeting advertisments based on a user's browsing behavior 
is known as behavioral targeting. 

\begin{figure}
\includegraphics[width=3in]{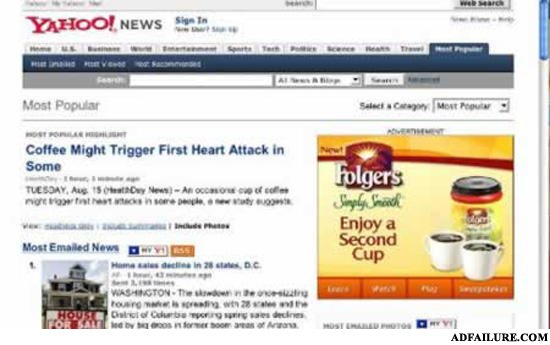}
\caption{Simple keyword targeting of ads can have unwanted effects}
\label{fig:coffee}
\end{figure}

How do ad networks know the context of the webpage?
Numerous approaches exist.
Some generate keywords based on the content of a webpage,
while others manually add keywords as meta data to 
indicate what the context of a web page is.
While conceptually simple these approaches are susceptible to
language use and 
can result in inappropriate and even macabre results as seen in 
Figure \ref{fig:coffee} \cite{coffeeadfail}.
Why this happens is due to the nature of keywords,
which can capture neither a user's interests nor intention.
There are many reasons why someone visits a webpage.
A user might read an article suggesting a link between coffee and
heart attacks because they drink a lot of coffee and have a weak heart.
Or their parents may have a heart problem, and she wants to determine
if this is something to forward to her parents. Or it was lunch time,
and the user was casually reading the news.
This last explanation is particularly troublesome now 
that many content providers optimize their headlines to maximize clicks.
Doing so creates a positive feedback loop where people's behavior 
is driven not so much by the content but by the optimized click-bait.
Therefore, even less intention can be inferred from the 
surfing behavior.

\subsection{Hashtags}
Another proxy for interests are hashtags in social media.
The idea is similar to keywords on webpages and suffers from the same issues.
It is even harder with hashtags as they are often 
associated with idiosyncratic memes and 
can be fragmented across many hashtags.
To combat this, companies like ESPN have attempted to standardize hashtag usage.
But this is an uphill battle as hashtags are inherently unstructured. 

\begin{wrapfigure}{r}{1.6in}
\begin{center}
\includegraphics[width=1.5in]{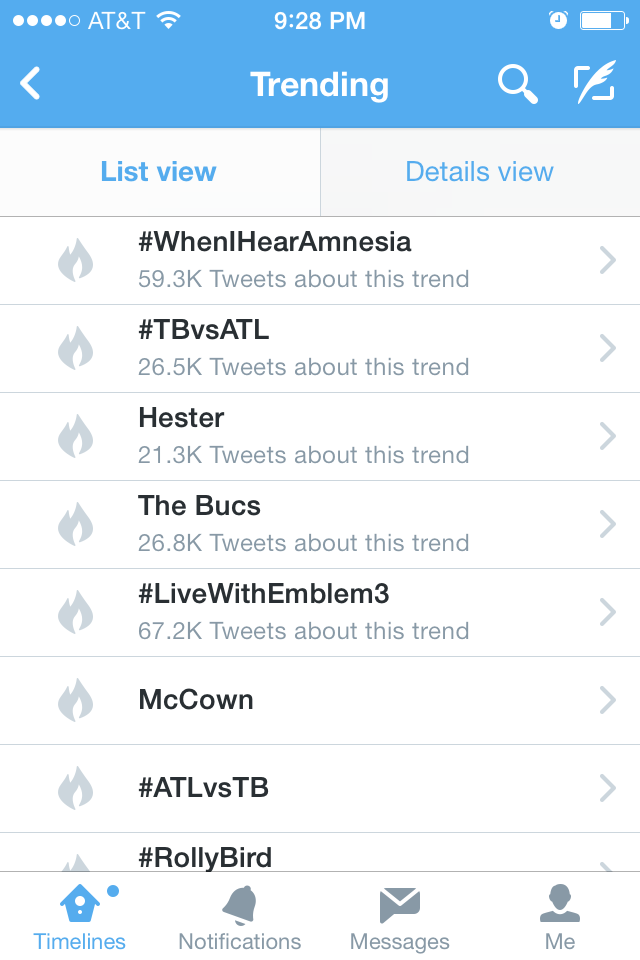}
\end{center}
\caption{The context and meaning of hashtags are not always clear}
\label{fig:twitter-trending}
\end{wrapfigure}

Even if you can standardize hashtags,
what exactly do they mean, and what drives their usage?
A casual look at trending hashtags on Twitter indicates that 
many focus on gossip and current events, 
as Figure \ref{fig:twitter-trending} shows. 
By definition, these are topics people are interested in,
but it's not clear how to transform this interest into meaningful engagement
based on the keyword alone.
Someone wanting to interact with users on \#WhenIHearAmnesia
would need to browse the associated timeline of tweets
to understand what it's about.
What is the next step? Is there a common thread besides the song
Amnesia that brings these people together? 
Knowing their interests would certainly help.

\begin{figure}[h]
\includegraphics[width=3in]{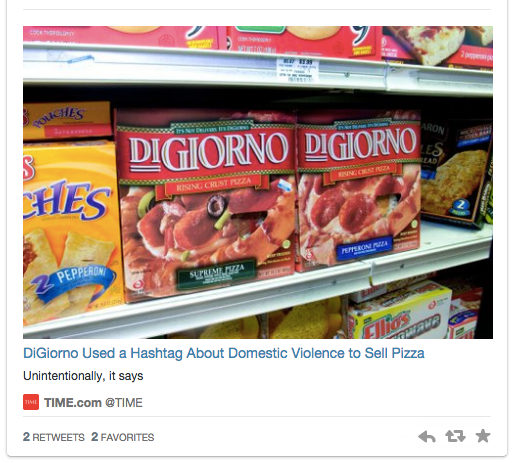}
\caption{Misunderstanding the context of a hashtag can result in negative
publicity}
\label{fig:digiorno}
\end{figure}

Not understanding an audience can be like walking through a minefield,
as many people are hostile to businesses exploiting hashtags.
These cavalier attempts at engagement tend to backfire,
creating a firestorm of negative publicity.
Figure \ref{fig:digiorno} highlights a situation where the person
managing the DiGiorno twitter account made a glib comment on
the \#WhyIStayed hashtag. Unfortunately, this tag was used to share
stories on domestic violence, which understandably angered many people
\cite{digiorno}.
This underscores the importance of moving beyond keyword targeting,
which is insufficient for establishing meaningful dialogue. 
An interest-based approach that recognizes and respects people’s passions 
is more likely to resonate with people instead of alienate them.

\subsection{Graph-based interests}
\begin{figure}
\includegraphics[width=4in]{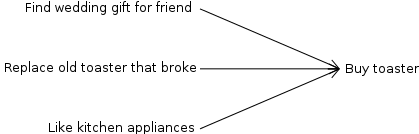}
\caption{Many interests can result in the same action making reverse inference unreliable}
\label{fig:interest-action}
\end{figure}

Behavioral targeting and collaborative filtering produce shallow
interests at best.
Deep interests are necessary for true outreach but are hard to identify.
The fundamental problem with these approaches 
is that causality is inverted. 
Our actions are driven by our interests,
and many interests can lead to the same action.
This convergence towards behavior means it's difficult to 
work backwards from there to the source intention or interest. 
Assuming that a purchase implies interest in a product or product category 
is too simplistic. Let's see how this assumption breaks down.
Suppose a friend is getting married, and you buy the couple a wedding present.
What is your interest in the present? If it is a toaster,
should a model infer that you are interested in cooking?
These multiple starting points are illustrated in Figure 
\ref{fig:interest-action}.
Attempting to infer interests based on purchases can therefore result in 
incorrect conclusions.
Rather than working backwards from effect to cause,
it is better to model interests directly.

Our social network provides a more direct and accurate picture 
of our interests than other methods. 
The idea is actually quite simple: we tend to connect with and 
follow people we are interested in. 
Someone who is passionate about cooking will probably 
follow chefs and people who tweet recipes. 
On the other hand, someone who buys a kitchen appliance as a wedding gift will most likely not follow those same people. 
Whom people follow is therefore directly linked with their interests.
Clearly people do not follow others if they have no interest in them. 
Not only that, but people tend to cluster together 
within communities of similar interests,
which is known as homophily \cite{2062393}.
In other words, when you follow one cook, 
it's likely that you'll follow other cooks 
(such as the cooks that that cook follows).
In this way, people form tight communities that 
share a common interest.
This clustering effect makes the model more resilient to noise 
than other methods. 

The beauty of this approach is that it recognizes the 
inherent multitude of interests people have.
Most people have more than one interest,
implying that they follow people in multiple communities.
An interesting and useful social phenomenon is that 
these groups are typically disconnected
\ref{fig:disparate-networks}. 
Hence, all of a user's interests can be inferred based on 
partitioning their social graph 
\footnote{At least to the extent that they express those interests on Twitter}.

\section{Graph Theory}
Social network analysis is part of a branch of mathematics 
known as graph theory, which studies how things are connected. 
Terminology can vary, but the essence is that a graph $G = (V,E)$ 
comprises a set of vertices (aka nodes) and edges (relationships) between them.
The number of edges radiating out from a vertex is its degree. 

As abstract mathematical structures, 
graphs can represent many things.
The beginning of graph theory is credited to Leonhard Euler who 
analyzed the ways to cross the seven bridges of Konigsberg.
In his analysis Euler abstracted the land masses as vertices and 
bridges as edges between the vertices.
For social networks, graphs represent the relationship between people.
In this model the graphs are typically directed,
because the direction of a connection matters:
if you follow someone, that person might not follow you back.
In Twitter, this distinction is known as friends (people a user follows)
and followers (people that follow the user).

\subsection{Community detection}
\begin{figure}
\includegraphics[width=\textwidth]{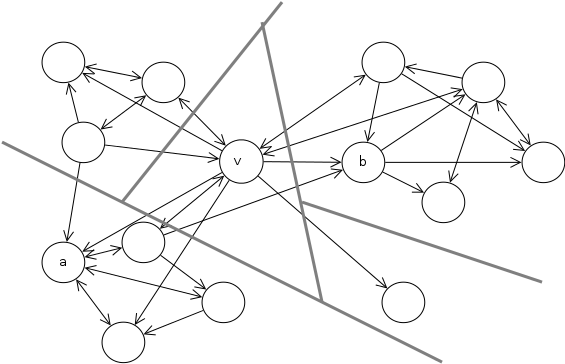}
\caption{Stylized network of disparate communities}
\label{fig:disparate-networks}
\end{figure}

As we get more involved with a given topic, 
we tend to follow more and more people within the same community.
This strengthens the bonds within that particular interest group.
Consequently, the connections within a particular interest group 
are greater than those between groups. 
This structure within a graph is called a community,
and finding such structures is 
is known as community detection. 
Numerous techniques exist for identifying communities within a graph. 
Communities can be constructed by splitting a graph into sub-graphs or 
merging sub-graphs into larger graphs.
Either way, these techniques tend to partition a graph so that 
the union of all communities contains all the nodes of the original graph
as illustrated in Figure \ref{fig:disparate-networks}.

The Girvan-Newman algorithm uses the betweenness measure 
to divide a graph into communities \cite{Girvan11062002}.
Vertex betweenness is a measure of network centrality,
indicating the magnitude of connections that a particular vertex has.
To understand how it works,
consider a vertex $v$ and two other vertices in the graph, say $a$ and $b$. 
For each pair of vertices, there will be a 
certain number of paths that can be drawn between them. 
Of these, only the shortest paths matter.
Some of these shortest paths will pass through $v$ while others will not. 
Let's call this value $s_{a,b}$.
Now imagine doing this for all nodes in the graph. 
The sum $\sum s_{a,b}, \forall a,b \in V, a \neq b \neq v$
is the vertex betweenness of $v$.
While this approach works reasonably well,
it is inefficient and is only practical with small networks. 

An alternate measure is modularity,
which is used in the Louvain method \cite{2010PhRvE..81d6106G}.
Modularity analyzes the edges of a network within a partitioned graph.
First count the number of edges within each community 
versus the number between communities.
Now construct a random graph using vertices with the 
same degree as the original graph.
Using the same partition, calculate the ratio of edges 
within the community versus between the communities.
Do this many times to get the expected value.
The difference between the actual number and the expected number is the modularity of the graph. Community detection thus becomes an optimization problem, where modularity is maximized for a network. What's nice about this approach is that there's a closed form solution for the expected value, which simplifies the computation. 
However, it is also known to have poor resolution and can miss small communities
\cite{2010PhRvE..81d6106G}.

An altogether different approach takes advantage of Monte Carlo simulation.
The so-called walktrap algorithm takes advantage of the insight that 
certain graphs will have more connections within communities than 
between communities \cite{walktrap}.
Therefore, taking a number of short random walks on each vertex will 
yield a set of vertices that are more likely to be within the same community.
With enough iterations, it is possible to construct a Markov Chain from all the random walks. A probability-based distance metric can then be defined which provides the necessary link with an agglomerative clustering method.

\section{Remarks and Conclusion}
By partitioning an individual's social graph, we construct the so-called interest map. 
The interest map displays the different interest groups and calculates descriptive text for each group.
Our research has revealed graphs that subdivide broad communities, such as alumni, into separate groups depending on their connections. This makes sense since people tend to have different circles of friends delineated by interests. A person's interest map not only reveals their college affiliation but also whether they follow their college football team. 
The interest map also detects family circles, which are separated from work circles. As someone changes jobs, multiple work circles form, each of which clearly indicate place and job function.

The biggest limitation to a graph-theoretic approach to interest detection is two-fold. Given person $a$ with interests $A$. 
The interests identified from their social graph $\hat{A} \subseteq A$. 
Ideally we'd like $\frac{|\hat{A}|}{|A|} \approx 1$, but this is intractable. 
In addition, since $A$ is an internal state, $|A|$ is unknowable meaning that 
$\frac{|\hat{A}|}{|A|}$ cannot be computed!
As an example, some people use Twitter purely for entertainment purposes. This type of person may only follow sports teams and/or celebrities. Suppose person $a$ has an interest map dominated by basketball and football. What other interests does this person have? Is it reasonable to assume that she has other interests? 

Though this presents a challenge, it is not as limiting as it seems. 
One way to think about it is in terms of precision and recall,
which is a popular way to measure the performance of 
machine learning algorithms \cite{Davis06therelationship}. 
Precision is defined as the number of true positives (TP)
versus total guesses, or $\frac{TP}{TP + FP}$, where FP are false positives. 
Recall, on the other hand, is the ratio of correct positive guesses to 
total positive cases, or $\frac{TP}{TP + FN}$, where FN are false negatives.
The best models have both high precision and recall,
but typically a model is tuned to give the optimal balance between the two.
An example is a spam filter,
which is optimized for precision at the expense of recall. 
The rationale is that the cost of a false positive 
(incorrectly guessing that an email is spam) is high,
whereas a false negative (incorrectly guessing spam is not spam) is low.
In our case, a false positive is incorrectly identifying someone's interests,
while a false negative is identifying no interest when one exists. 
Here again, it is more important to have good precision than recall. 
Even if we don't know all the interests of someone, so long as the one's we identify are correct, the model is useful. 
Notice that our definition of recall is actually 
a reformulation of the proportion of identified interests 
$\frac{|\hat{A}|}{|A|}$. 
In other words, we cannot compute the recall of the model,
but we also know that it is unnecessary. 

Interest maps are a powerful tool for identifying people's interests.
They provide a degree of intimacy that can lead to 
meaningful relationships with others. 
Understanding an individuals true interests provides opportunities 
for authentic engagement. Classical approaches for inferring someone's 
interests assumes that shopping or surfing behavior is a 
reasonable proxy for interests. 
This assumption suffers from two major issues.
First, causality is inverted between interests and actions.
Second, these proxies are susceptible to noise.
A more resilient approach is to analyze an individuals social network,
which more accurately represents their interests.
Organizations can connect with their audience on a deeper, more personal way.
Doing so not only is more efficient but can build lasting, loyal relationships.

\bibliographystyle{plain}
\bibliography{biblio}

\end{document}